\documentclass[12pt, letterpaper]{article}
\usepackage[utf8]{inputenc}

 \usepackage{setspace} 

\usepackage{graphicx}

\usepackage{amsmath, amssymb}
\usepackage{xcolor}
\usepackage{algpascal}
\usepackage{algc}
\usepackage{algorithm} 
\usepackage{algpseudocode}
\usepackage{slashbox}
\usepackage[hyphens]{url}

\title{Robust Android Malware Detection System against Adversarial Attacks using Q-Learning}

\author{
Hemant Rathore, Sanjay K. Sahay, Piyush Nikam\\
BITS Pilani, Department of CS \& IS, Goa Campus\\
\texttt{\{hemantr, ssahay, h20180057\}@goa.bits-pilani.ac.in}
\and
Mohit Sewak\\
Security \& Compliance Research, Microsoft, India\\
\texttt{mohit.sewak@microsoft.com}
}

\date{} 

\begin{document}
\maketitle

\begin{abstract}
Since the inception of Android OS, smartphones sales have been growing exponentially, and today it enjoys the monopoly in the smartphone marketplace. The widespread adoption of Android smartphones has drawn the attention of malware designers, which threatens the Android ecosystem. The current state-of-the-art Android malware detection systems are based on machine learning and deep learning models. Despite having superior performance, these models are susceptible to adversarial attack. Therefore in this paper, we developed eight Android malware detection models based on machine learning and deep neural network and investigated their robustness against the adversarial attacks. For the purpose, we created new variants of malware using Reinforcement Learning, which will be misclassified as benign by the existing Android malware detection models. We propose two novel attack strategies, namely single policy attack and multiple policy attack using reinforcement learning for white-box and grey-box scenario respectively. Putting ourselves in adversary’ shoes, we designed adversarial attacks on the detection models with the goal of maximising fooling rate, while making minimum modifications to the Android application and ensuring that the app's functionality and behaviour does not change. We achieved an average fooling rate of $44.21\%$ and $53.20\%$ across all the eight detection models with maximum five modifications using a single policy attack and multiple policy attack, respectively. The highest fooling rate of $86.09\%$ with five changes was attained against the decision tree based model using the multiple policy approach. Finally, we propose an adversarial defence strategy which reduces the average fooling rate by threefold to $15.22\%$ against a single policy attack, thereby increasing the robustness of the detection models i.e. the proposed model can effectively detect variants (metamorphic) of malware. The experimental analysis shows that our proposed Android malware detection system using reinforcement learning is more robust against adversarial attacks.


\end{abstract}

\section{Introduction}\label{intro}
Android smartphones have evolved tremendously in the last decade, and today have reached more than half of the world's population \cite{globaldr}. Annual Android smartphone sales are expected to reach $1.32$ billion in $2020$ \cite{marketshare}. The broad acceptance of Android is due to its open-source nature, robust development framework, multiple app marketplaces, large app stores, etc \cite{tam2017evolution}. Growth of Android OS is also fueled by recent development of $4G$ and $5G$ internet technologies. Internet is currently the primary attack vector used by malware designers to attack the Android ecosystem \cite{ye2017survey}.

Malware (\textit{Mal}icious Soft\textit{ware}) is any software program designed and developed with an evil intent against any target \cite{ye2017survey} \cite{faruki2014android}. Malware is not new to the computing environment. The first malware \textit{Creeper}\footnote{\url{https://enterprise.comodo.com/malware-description.php}} was a self-replicating program written by Bob Thomas in 1971 for the TENEX operating system \cite{ye2017survey}. On the other hand, the first malware on the Android platform was a trojan named \textit{Trojan-SMS.AndroidOS.FakePlayer.a}\footnote{\url{https://www.f-secure.com/v-descs/trojan_android_fakeplayer.shtml}} detected in August 2010 which sends SMS messages to premium-rate numbers without the user's knowledge \cite{ye2017survey}. A report by AV-Test suggests detection of $65.26$ million malware on the Android ecosystem in $2011$. Since then, there has been an exponential growth of malicious applications with a current estimate of $1049.7$ million for 2020 \cite{avtest}. According to GDATA antivirus report, more than $10,000$ new malicious apps are detected every day and also malware designers are uploading an infected Android app every eight seconds \cite{gdata}. ISTR by Symantec reported a 33\% increase in mobile ransomware with one out of every thirty-six mobile devices had the high-risk app installed on it \cite{symc}.

The primary defence against any malware attack is provided by the anti-malware research community and the anti-virus industry (Bitdefender, Kaspersky, McAfee, Symantec, Windows Defender etc)\footnote{\url{https://attackevals.mitre.org/}}. Currently, most of the anti-viruses work on the signature, heuristic and behaviour-based detection engines \cite{ye2017survey} \cite {sahay2020evolution} \cite{tam2017evolution}. These mechanisms are excessively human dependent, time-consuming and not scalable and thus cannot detect next-generation polymorphic/metamorphic malware \cite{ye2017survey} \cite{tam2017evolution}. Recently machine learning and deep learning have shown promising results in various domains like image recognition, natural language processing, recommendation systems etc. Thus, the anti-malware research community have also explored malware detection using machine learning and achieved promising results \cite{ye2017survey} \cite{tam2017evolution} \cite{rathore2018malware} \cite{faruki2014android}.

Classification models based on machine learning and deep learning are susceptible to adversarial attacks. Good fellow et al. in 2015 performed an adversarial attack on an image classification model. They showed that small but intentionally worst-case perturbations could result in the classification model misclassifying the test image.  Also, sometimes these misclassifications are achieved with higher confidence \cite{goodfellow2014explaining}. Earlier, non-linearity and overfitting were used to define the robustness of classification models which is not the case today \cite{goodfellow2014explaining}. This fact can be exploited by malware designers to perform targeted and non-targeted attacks using adversarial learning.

Reinforcement learning (RL) can be used to generate intentional perturbations which can be visualized as a min-max game where a reinforcement learning agent determines a sequence of actions to maximize the return based on the reward function \cite{fonteneau2010towards}. Adversarial attacks are highly dependent on the attacker's knowledge about the malware detection system. This knowledge can consist of information about the training dataset, features information, model architecture, and classification algorithm used to construct the models\cite{papernot2016limitations} \cite{serban2018adversarial}. If the adversary has complete knowledge about the malware detection system, then it is known as a white-box scenario \cite{serban2018adversarial}. On the other hand, if the adversary attacks without any prior knowledge about the system, then it is called as black-box scenario \cite{serban2018adversarial}. However, the above cases are two extreme ends of the curve for any real-world scenario. Thus we started with white-box attack setting on malware detection models. Later, we also performed a grey-box attack where an adversary has limited knowledge about the malware detection system, which does not include any information about the model's architecture or classification algorithm.

Therefore in this paper, we proposed a novel adversarial attack for the white-box scenario, namely single-policy attack as it crafts perturbations governed by a single policy extracted from a Q-table. The optimal policy governs the modification attack, where the feature vector extracted from a malicious Android application is modified to be misclassified as benign by different Android malware detection models built using classical, ensemble and deep learning algorithms. The goal of the optimal policy is to modify the maximum number of malicious applications and generate new variants which are misclassified by detection models. Also, modifications are minimized in each application to reduce the overall cost of the attack. These modifications are syntactically possible and do not disrupt the behavioural or functional aspect of the Android applications. We also proposed a multi-policy attack for the grey-box scenario, which consists of a set of optimal policies extracted from many Q-tables and used parallelly for adversarial attacks. To the best of our knowledge, this is the first investigation of adversarial attacks on the grey-box scenario in the area of malware detection.

In summary, we have made the following contributions through this paper for the development of the robust Android malware detection system against adversarial attacks:

\begin{itemize}
  \item We proposed two novel adversarial attacks on Android malware detection system followed by a defence mechanism to improve the overall robustness of the detection models.
  \item We designed a single-policy adversarial attack for the white-box scenario, which achieved an average fooling rate of $44.28\%$ with a maximum of five modifications across eight diverse set of detection models.
  \item We also devised a multi-policy attack for the grey-box scenario, which obtained an average fooling rate of $53.20\%$ with a maximum of five modifications across the same eight detection models.
  \item Finally, we proposed a defence mechanism which reduces average fooling rate by threefold and twofold against single policy and multiple policy attack respectively to improve the overall robustness of the Android malware detection system.  
\end{itemize}

The remainder of the paper is organized as follows: Section 2 starts with a brief discussion on problem definition and proposes an architecture to construct a robust Android malware detection system. Further in the section, the design of adversarial attacks on malware detection models is explained, followed by defence strategies. Section 3 discusses the experimental evaluation, which contains details of the experimental setup like classification algorithms, dataset, feature evaluation, and performance metrics. Later, experimental results of adversarial attacks using the single policy and multiple policy attack are explained followed by results using defence strategy. Section 4 contains the related work in the domain. Finally, Section 5 concludes the paper and presents the future direction of work.

\section{Problem Overview and Proposed Architecture}

In this section, we will first explain the problem definition in sub-section \ref{problem_def}. Sub-section \ref{proposed_arch} will introduce the proposed architecture for the construction of a robust Android malware detection system against adversarial attacks. Also, figure \ref{framework} briefly explains different components of the proposed architecture. Later, sub-section \ref{devop_aa} will explain the development of the adversarial attack strategy followed by a discussion on the adversarial attacks (sub-section \ref{aa_theory}) and defence (sub-section \ref{defense_theory}) on Android malware detection system.

\subsection{Problem Definition}\label{problem_def}



To design a system against adversarial attack on the Android malware detection system, let us consider the dataset:

\begin{equation}
D = \{(x_i, y_i)\} \in (X,Y)
\end{equation}


\noindent where $D$ consists of $m$ malicious and $b$ benign apps ($\vert b \vert \approx \vert m \vert$). Here $X$ and $Y$ represents a set of Android applications and their corresponding class label (malware or benign). If an application is benign, then $y_i$ is set to $0$ otherwise $1$. $x_i$ and $y_i$ represents the $i^{th}$ Android application, and its class label respectively. Using the dataset $D$ different classification algorithms can be trained to build an effective, efficient and robust Android malware detection model.


In the malware evasion attack, the goal is to modify the maximum number of malicious applications $x_i$  such that they are misclassified as benign by the Android malware detection model. The proposed modifications in the set of malicous applications:

\begin{equation}
M = \{(x_i, y_i)\}, \quad  \forall \; y_i \in M, y_i = 1
\end{equation}

\noindent should be syntactically possible to integrate with the $i^{th}$ application. Also, these modifications should be possible to incorporate in the Android application without any behavioural or functional changes in the application. Another goal should be to minimize the modifications in $x_i$ to reduce the overall cost of evasion attack. Let us assume a large number of misclassified malicious samples $M’$ are successfully constructed from $M$ using evasion attack. An adversary can use these $M’$ malicious Android applications to fool the detection models and reduce the overall accuracy of any real-world Android malware detection model. However, these $M’$ samples can also be used by the defender to prepare a defence against evasion attacks.

\subsection{Architecture overview of the proposed method}\label{proposed_arch}

\begin{figure}[htbp!]
	\centering
	\includegraphics[width=1.0\linewidth]{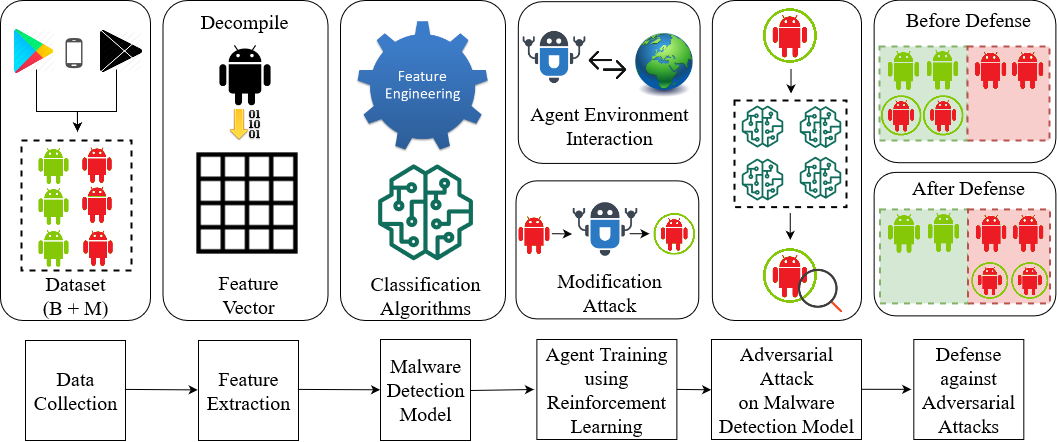}
	\centering
	\caption{Overview of the proposed architecture for adversarial attack and defence}
	\label{framework}
\end{figure}

Figure \ref{framework} illustrates the proposed approach to construct a robust Android malware detection system. Firstly, Android applications (malware and benign) were gathered from various authentic sources. These applications were decompiled using an open-source tool. Then a parser was developed to scan through each decompiled application and construct the feature vector.  We extracted Android permissions from all the downloaded applications and populated them in the feature vector.  Since the feature vector suffers from the curse of dimensionality thus we performed a detailed analysis of the permission vector using different feature engineering techniques. Then we used different classification algorithms to construct eight distinct malware detection models and used them as the baseline. In the next phase, we designed adversarial attack strategies on the baseline detection models using reinforcement learning. The idea is to modify the malicious Android application such that the detection model(s) misclassifies it as benign.  Reinforcement learning agents were allowed to interact with the environment consisting of the malicious applications and detection models. Learnings from these interactions were used to define an optimal policy for the adversarial attacks. These attack strategies could be used by malware designers or adversaries to successfully modify malicious Android applications and force any real-time antivirus to misclassify them as benign. As an anti-malware designer, we should be able to defend against these type of attacks, and thus we propose the defence strategy. Our analysis shows that the defence strategy improved the robustness of existing malware detection models and reduced the chance of zero-day attacks.

\subsection{Development of Adversarial Attack Strategy} \label{devop_aa}

Reinforcement Learning has gained tremendous interest in the past few years due to its ability to perform particular tasks without specifying explicit steps \cite{luong2019applications}. The agent is trained based on rewards and penalties to perform a given task. The trial and error learning methodology of an RL agent consists of two main strategies. The first strategy is to find the optimal action in action-space in the given environment \cite{kaelbling1996reinforcement}. The second strategy is to estimate the utility of the actions using dynamic programming \& statistical learning and then to choose the optimal action \cite{kaelbling1996reinforcement}. Another area in machine learning that has gained interest recently is adversarial learning where the ML models are trained by intentionally injecting adversarial samples. In this, Goodfellow et al. demonstrated that adversarially trained models show greater robustness against the adversarial attacks \cite{goodfellow2014explaining}.

To simulate an attack using RL on Android malware detection models, we defined an environment, set of states, and set of possible actions. The RL simulation environment can be formalized using the Markov Decision Process (MDP), which is a mathematical framework for modelling the decision-making process. It consists of a finite set of malware detection models $C \:(C = \{c_1, c_2 .... c_M\})$, a finite set of states $S\:(S = \{s_1,s_2 .... s_N\})$, a finite set of actions $A\:(A = \{a_1, a_2 .... a_k\})$, a reward function $R$ and a policy $\pi$. The proposed environment consists of Android malware detection models along with the application permission vectors as states. The permissions extracted from the Android applications are used to define the finite set of possible states. A set of models $C$ consists of malware detection models, and the set of actions $A$ is defined with the help of permissions $P \:(P = \{p_1, p_2 .... p_k\})$. The notations and symbols used in the proposed work are mentioned in Table \ref{MDPdefine}.

\begin{table}[ht]
\caption{Notation and symbols used in the paper}
\begin{center}
\begin{tabular}{|c|l|}
\hline
\textbf{Name} & \multicolumn{1}{c|}{\textbf{Description}}                                                                                                      \\ \hline
$C$             & set of Android malware detection models                                                                                                    \\ \hline
$S$             & \begin{tabular}[c]{@{}l@{}}set of states representing permission vector extracted\\ from malicious and benign Android application\end{tabular} \\ \hline
$A$             & set of actions                                                                                                                                 \\ \hline
$R$             & reward function                                                                                                                                \\ \hline
$\pi$       & mapping from states to actions                                                                                                                 \\ \hline
\end{tabular}
\end{center}
\label{MDPdefine}
\end{table}

During the training phase, the agent starts interacting with the environment which is starting from state $s_0$ and performs transitions with the help of actions that are governed by an optimal policy $\pi^*$ to finally reach the goal state $s_g$. The collection of these transitions taken over discrete time steps constitutes an episode $e_t$. The resultant value of an episode $e_t$ depends on the actions taken at every time step and is defined by the state-value function given as:

\begin{equation}
V^\pi(s) = E_\pi[r_{t+1} + \gamma * r_{t+2} + ... | s_t = s]    
\label{state-value function}
\end{equation}

The state-value function is used to define the expected value obtained starting at state $s$ and following the policy $\pi$ with a discount factor $\gamma \in [0,1)$ for successive transitions. $E_\pi[\;]$ represents the expectation over reward obtained using policy $\pi$. At any step, the state $s_t$ is represented by a permission vector and action $a_t$ is represented by the modified permission $P_i$ ( $i \in $ [$0$, size of permission set] ). Similar to state-value function, the action-value function is defined as:

\begin{equation}
Q^\pi(s,a) = E_\pi[R_\pi | s+t = s, a_t = a ] 
\label{action-value function}
\end{equation}

\noindent which determines the expected value obtained by taking action $a_t$ from state $s_t$ following the policy $\pi$ for all the successive transitions. In the proposed work, action-value function represents the likelihood of modifying permission $p_i$ in the given feature vector to maximize the probability of misclassifying the malicious application as benign while minimizing the modification cost. The modification cost signifies the number of permissions modified to reach the goal state $s_g$ from $s_0$. A state $s_t$ is marked as $s_g$ once the benign probability of the samples exceeds $0.5$ as discussed in following equation:


\begin{equation}
 P(x_i) =\begin{cases}s_t, & P_b > 0.5  \:(\text{classified as benign})\\s_g, & P_b \leq 0.5 \:(\text{classified as malware})\end{cases}
 \label{pandmvalue}
\end{equation}

The reward of an agent is governed using the reward function:


\begin{equation}
r = w_1 * P_b - w_2 * N_m + w_3 * S_g
\end{equation}

\noindent where $P_b$ is the benign probability, $N_m$ is the number of modifications, and $S_g$ represents status of the goal state. $w_1$, $w_2$, and $w_3$ are weights associated with $P_b$, $N_m$, and $S_g$, respectively.

Based on the reward function, the agent is rewarded to increase the benignness of the malicious sample and to reach the goal state $s_g$. Also, it is penalized for every modification to reach the goal state. Finally, the optimal policy:

\begin{equation}
\pi^* = \text{argmax}_\pi V^\pi(s), \quad  \forall s  \in S
\label{eq_optimalpolicy}
\end{equation}

\noindent is calculated based on an agent's interaction with the environment $\epsilon$ over several million episodes. The starting state $s_0$ of an episode $e_t = (s_t, a_t, r_t, s_{t+1})$ is selected randomly. An episode $e_t$ is terminated once the agent reaches the goal state $s_g$ which in our case is to modify the malware sample to be misclassified as benign. The agent-environment interactions are stored in a memory $D$ called replay buffer. A Q-table (table \ref{sampleqtable}) is created once the size of the replay buffer is large enough to obtain a policy that is close to an optimal policy $\pi^*$.

\begin{table}[ht]
\caption{Sample Q-table}
\begin{center}
\begin{tabular}{|c|c|c|c|c|c|}
\hline
\backslashbox{State}{Permission}
        & $P_0$ & $P_1$ & .. & .. & $P_n$ \\ \hline
0 .... 0 &    &    &    &    &    \\ \hline
  ....   &    &    &    &    &    \\ \hline
  ....   &    &    &    &    &    \\ \hline
1 .... 1  &    &    &    &    &    \\ \hline
\end{tabular}
\end{center}
\label{sampleqtable}
\end{table}

The row of the Q-table represents the states, i.e. the application permission vector and columns represent the Android permission $P_i$ to be modified. The Q-table contains $n+1$ columns where the entries in $i^{th}$ column correspond to the expected value to be obtained after modifying the $i^{th}$ permission. The highest value in $0^{th}$ column denotes that the sample is benign and no further modification is necessary. The agent is allowed to modify one permission per time step. An interaction represents a transition from one state $s_t$ to $s_{t+1}$ by performing an action $a_t$ taken according to $\pi^*$. The reward function is created in a way to maximize the chance of modifying a malware application to be misclassified as benign with the least number of alterations. Monte-Carlo Every-Visit approach \cite{sutton2018reinforcement} is used to populate the values of the Q-table. Once the Q-table is constructed, it is used to extract the optimal policy $\pi^*$ to modify the permission vector. A policy $\pi$ could be visualized as a lookup table consisting of an optimal action $a_t^*$ at every given state. These actions maximize the reward of an agent and the probability to reach the goal state.

\subsection{Adversarial Attack on Android Malware Detection Model} \label{aa_theory}

\begin{figure}[htbp!]
	\centering
	\includegraphics[width=\linewidth]{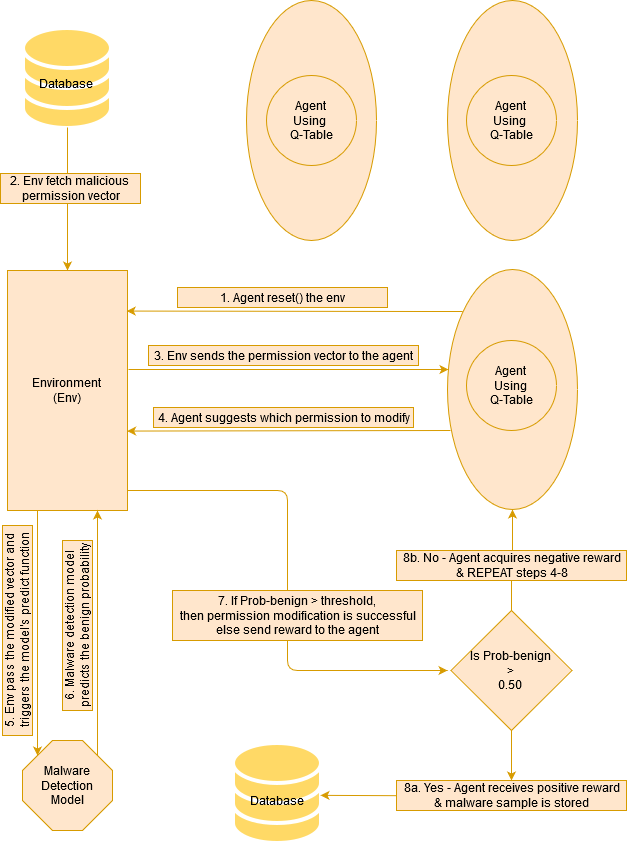}
	\centering
	\caption{Adversarial attack on Android Malware Detection system}
	\label{AgentFlow}
\end{figure}

Figure \ref{AgentFlow} depicts the eight-step process-flow of the adversarial attack strategy on the Android malware detection models. The agent starts by resetting the environment $\epsilon$ and fetching the permission vector from the database which is then sends to the agent. After receiving the feature vector, the agent manipulates the permission $P_i$ of the malicious sample based on the current optimal policy ($\pi^*$). Multiple agents can be used during adversarial attack to simulate multi- policy attack scenario. The agent returns the modified vector to the environment which further passes it to the Android malware detection model and triggers the model’s predict function. The detection model performs classification and returns the benign probability to the environment. This benign probability is used to calculate the reward $r_t$ using the reward function. The environment sends this reward and the next state $s_{t+1}$ to the agent. If the benign probability is greater than $0.5$, the process is terminated as the sample is benign. It also suggests that the agent is successfully able to modify the malicious application to be misclassified as benign by the detection model. If the benign probability is less than or equal to $0.5$ it suggest that the sample is still malicious, and a few more modifications are required to convert it to a form that is misclassified as benign. This iterative process supported by reward function helps the agent to reach the goal state.

\paragraph{\textbf{White-Box Attack Scenario:}}
As discussed in section \ref{intro}, an adversary can attack an Android malware detection system in two different scenarios. In a white-box attack scenario, we assume that the adversary has complete knowledge about the dataset,  feature information, model architecture and the classification algorithm. For a white-box scenario, we designed a Single Policy Attack (SPA) using a single Q-table where only one optimal policy $\pi^*$ is used to modify the permission vector of a given malware sample. The permission modification process continues until the malicious sample is misclassified as benign by the detection model. The limitation of this approach is that it assumes that adversary’s has complete knowledge about the detection system, which is partially incorrect for any real-world scenario.

\paragraph{\textbf{Grey-Box Attack Scenario:}}
In this scenario, we assume that the adversary has only limited knowledge which does not include any information about model architecture and classification algorithm used to build the detection model. Thus for grey-box setting, we designed a Multiple Policy Attack (MPA) using many Q-tables consisting of a set of optimal policies $\pi_i^* \in \prod$ (set of optimal policies) in contrast to the single policy in SPA. The set of optimal policies is obtained from different Q-tables created using a set of replay buffers storing environment-agent interaction of a set of agents. These policies are parallelly used to modify the malware sample. The modification process terminates as soon as the sample is misclassified as benign using any of the optimal policy $\pi^*$. These benign-looking samples are added to a set of new malware samples obtained during the SPA and MPA. Attacks using multiple Q-tables help in formalizing grey-box attack scenarios where adversaries can perform successful attacks even without complete knowledge of the Android malware detection system. 

The algorithm \ref{alg1} describes the MPA strategy against malware detection models. It accepts $X$, $Y$, $M$, \& $C$ as input which represent malicious and benign data, class labels, set of malware samples, and set of malware detection models respectively. The algorithm produces the following outputs: the modified malicious samples ($M^{'}$) which are misclassified as benign, the number of modified permissions($N_c$) and ID of the malware detection model. After necessary initializations, the first loop chooses a malware sample $d \in M$ from the database, and the second loop selects the detection model $c \in C$. The third loop depicts an agent with multiple optimal policies obtained from multiple Q-table. The permission vector is extracted using $F_{ext}$ (extraction function) and is vectorized as state $s$. Given a malicious sample $s$, an optimal action $a_t^*$ is chosen using the optimal policy $\pi^*$. A modification function $F_m$ is used to modify specific permission using the optimal action. The modified state $s_{modified}$ is used to trigger the predict function of the malware detection model $C_p$ which returns the benign probability. The attack procedure is terminated as soon as the benign probability $> 0.5$ is achieved. Finally, the malware sample is successfully modified to be misclassified as benign, and the tuple  ($s_g$,$n_c$,$c$) is added to the database where $s_g$ represents the new malware sample (goal state), $n_c$ represents the number of steps taken to convert the sample, and $c$ represents the ID of malware detection model.

\begin{algorithm}[ht]
	\caption{Algorithm for multiple Q-Attack}\label{alg1}
	\textbf{Input} : \\
	$\mathbf{X}$: permission vector extracted from Android application \\
	$\mathbf{Y}$: class label (0 depicts benign and 1 depicts malware) \\
	$\mathbf{M}$: set of malicious sample \\
	$\mathbf{C}$: set of malware detection models \\
	\textbf{Function} : \\
	$\mathbf{C_P}$: prediction function in classification models \\
	$\mathbf{F_{ext}}$: feature vector extraction function\\
	$\mathbf{F_m}$: feature vector modification function \\
	$\mathbf{P}$: permission to be modified \\	
	\textbf{Output} : ${Sample (s_g, n_c, c)}$\\
	$\mathbf{s_g}$: modified permission vector ((goal state))\\
	$\mathbf{n_c}$: number to permission modified\\
	$\mathbf{c}$: classification model
	\begin{algorithmic}[1]
		\For {$\text{each malware sample}$ $d \in M$}
	    	\For {$\text{each malware detection model}$ $c \in C$}
	    		   	\For {$\text{each policy}$ $\pi^*$ $\textbf{parallel}$}
                    	\State ${s \leftarrow F_{ext}(d)}$
                    	\State ${p \leftarrow \pi^*(s)}$            		
                    	\State ${s_{modified} \leftarrow F_m(s, p)}$
                    	\State ${P_b, n_c \leftarrow C_p(c, s_{modified})}$   
                    	\If {${(P_b \geq 0.5)}$}
                    	\hspace{17em}\smash{$\left.\rule{0pt}{5.5\baselineskip}\right\}\ \mbox{in parallel}$}
                    	    \State ${s_g \leftarrow s_{modified}}$
                    	    \State ${Sample (s_g, n_c, c) \leftarrow (s_g, n_c, c)}$   
                    	\Else
                    	    \State repeat steps $2$ - $5$ using $s_{modified}$
                    	\EndIf
            		\EndFor
			\EndFor
		\EndFor
	\end{algorithmic} 
\end{algorithm}

\subsection{Defence strategy against adversarial attacks} \label{defense_theory}

While evaluating the performance of the Android malware detection models with SPA and MPA, we found many new variants of malicious samples that are misclassified as benign. These adversarial malicious samples can be used by cyber-criminals to bypass the malware detectors to compromise the Android OS and steal sensitive information. Kurakin et al. proposed that injecting adversarial samples while retraining the existing detection models significantly improved the robustness against known and zero-day attacks \cite{kurakin2016adversarial}. Chinavle et al. used mutual aggregation based dynamic adversarial retraining against adversarial attack on spam detection models \cite{chinavle2009ensembles}. Ji et al. in DEEPARMOUR also found that adversarial retraining help in improving the robustness of the detection models based on random forest, multi-layer perceptron and structure2vec \cite{ji2019securing}. As a proactive strategy, in the proposed work, we retrained the existing detection models with these newly discovered variants of malicious samples. After completing the retraining process, the detection models are put into a test against both SPA and MPA.

\begin{figure}[htbp!]
	\centering
	\includegraphics[width=\linewidth]{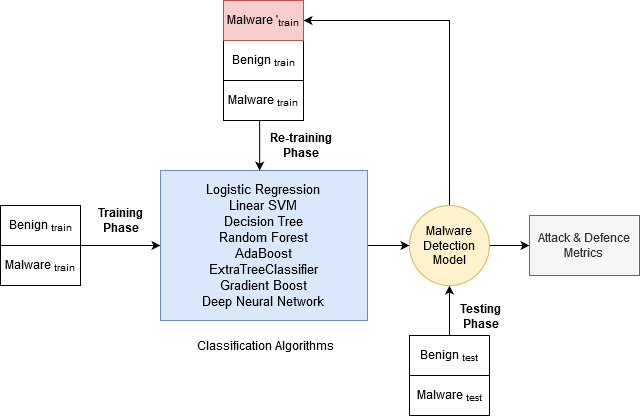}
	\centering
	\caption{Defence against adversarial attacks}
	\label{Defence}
\end{figure}

Figure \ref{Defence} shows the proposed defence strategy against adversarial attacks. Initially, we split the dataset into a train set (used for training the malware detection models) and test set (used to evaluate the performance of the detection model). During the adversarial attack, a large number of malicious samples fooled the detection model(s) and are misclassified as benign. These new variants of malicious samples with the corrected label are used to retrain all the detection models. The newly developed models are again subjected to SPA and MPA, and the performance is re-evaluated.

\section{Experimental Evaluation}

In this section, we discuss the experimental setup and results obtained from the SPA and MPA strategies on various malware detection models followed by the defence strategy results under different scenarios.

\subsection{Experimental Setup}

In the following sub-section, we first explain the set of classification algorithms used for the experiments followed by a description of the dataset and performance metrics.

\subsubsection{Classification Algorithms}

We use a diverse set of classification algorithms to validate the proposed approach. The set contains traditional machine learning algorithms (Logistic Regression (LR), Linear Support Vector Machine (SVM), Decision Tree (DT)), bagging algorithms (Random Forest (RF), ExtraTreesClassifier (ET)), boosting algorithms (Adaptive Boosting (AB), Gradient Boosting (GB)) and Deep Learning (Deep Neural Network (DNN)). We use k-fold validation for better generalization with the value of k set to $5$ for all the experiments. The SVM model is trained using the linear kernel, squared hinge as the loss function and l2 as the norm for the penalty. Since the dataset contains almost an equal number of malicious and benign samples, so the class weight is set to balanced. The DT based models are constructed using Gini impurity criteria with no limitation on the maximum depth of the tree and the minimum number of samples required at the leaf node. In LR models, we use lbfgs solver with l2 as the penalty norm. Again the class weight is set to balanced since the dataset contains equal number of malware and benign Android apps. RF models contained forest of $100$ trees with Gini impurity as splitting criteria. Again there is no limitation on the maximum depth and minimum samples for each decision tree in the forest. AB models use DT based boosting with SAMME.R as the boosting algorithm. The number of trees in AB is set to $100$. Again GB models are based on tree-based boosting with deviance as the loss function, and friedman mean square error as the split criteria. The number of trees at each boosting stage is set to $100$ while max depth of each tree is set to three for better generalization and to avoid overfitting. ET uses $10$ randomized DTs with Gini impurity as the split criteria. In each DT, the maximum depth and minimum samples to split is set to $0$ and $2$ respectively for better predictive accuracy and to control overfitting. The DNN model is a single-layer perceptron with one hidden layer containing $100$ neurons. It uses ReLU as the activation function with a batch size of $200$. The model is trained using Adam weight optimizer with a constant learning rate. We use all the above classification algorithms for building eight different malware detection models. In the adversarial attack phase, each of these models is attacked using different policies. Finally using the defence strategy, each of the models is made more robust against adversarial attacks.

\subsubsection{Dataset}

We have conducted all the experiments using the dataset of Android applications containing both malware and benign samples, as detailed below:

\paragraph{\textbf{Malware Dataset:}} Arp et al. downloaded $123,453$ Android applications from Google Play store \cite{googleplay} \& other third party alternative app distributors and found many malicious apps among them \cite{arp2014drebin}. They collected all the malware applications and named it the \textit{Drebin Dataset} which contains $5,569$ malware samples from more than $20$ malware families like FakeInstaller, DroidKungFu, Plankton, Opfake, GingerMaster, BaseBridge, Iconosys, Kmin etc. Drebin also contains all the samples from the Android Malware Genome Project \cite{zhou2012dissecting}. This benchmark Drebin dataset \cite{arp2014drebin} has been used by many researchers for developing Android malware detection system \cite{ye2017survey} \cite{faruki2014android} \cite{tam2017evolution}. 

\paragraph{\textbf{Benign Dataset:}} Google Play Store \cite{googleplay} is the official app store for distributing applications for the Android operating system. We downloaded $8500$ apps from the Play Store. In the past, many Android applications posted on the Google Play Store were found to be malicious and were taken out from the platform by Google. Thus we scan all the downloaded applications using VirusTotal \cite{virustotal} (which is a subsidiary of Google which aggregate the results from more than $50$ antivirus engines). We labelled an Android application as benign if none of the antivirus engines from VirusTotal declare it as malicious. Finally, we remove the malicious samples from the dataset and labelled all the remaining apps ($5,721 $) as benign.

\subsubsection{Feature Extraction}\label{Feature Extraction}

Android is a mobile operating system built on the modified Linux kernel. Its security architecture is designed with the Android permission system built in the centre to protect the privacy of the user. An app must request permission from the Android system to access data (call log, SMS etc.) or system resource (WiFi, Bluetooth etc.). We use Android permissions to build our baseline, attack and defence models. We perform feature extraction of Android permissions using Apktool \cite{apktool} by decompiling all the applications and then analyzing AndroidManifest.xml file. We develop a parser to prepare a master permission list containing all the $197$ Android permissions. This list is validated with both the official Android documentation and Android Studio compiler. Another parser is developed to scan through each application and populate the feature vector based on requested permission by that application. We execute the parser on the complete dataset containing both malicious and benign samples. The final malware and benign feature vectors is $5,560 \times 197$ and $5,721 \times 197$, respectively, where a row represents an Android application, and column denotes permission. An important point to note here is modifying the Android permissions in an application does not disturb the syntactical, behavioural and functional properties of the application. Thus the applications can be successfully recompiled with the modified permissions.

\subsubsection{Performance Metrics}

We use the following performance metrics collected from different malware detection models to evaluate the experimental results:

\begin{itemize}
  \item\textit{True Positive} (TP) is the outcome when predicted, and the actual value is True. It signifies that malware prediction by the detection model is correct.

    \item\textit{False Positive} (FP) indicates a classification error when the predicted value is True, but the actual value is False. It signifies that the detection model has wrongly classified the benign sample as malicious.

    \item\textit{True Negative} (TN) is the outcome when predicted, and the actual value is False. It also signifies that benign prediction by detection model is correct.

    \item\textit{False Negative} (FN) indicates a classification error when the predicted value is False, but the actual value is True.  It signifies that the detection model has wrongly classified the malware sample as benign.

    \item\textit{Accuracy} (Acc.) is the ratio of correct predictions and the total number of the predictions. It denotes the number of malicious samples classified as malware and benign samples as benign, divided by the total number of predictions. A higher value of accuracy signifies that malware detection is correctly predicting the labels with high confidence. 

\begin{equation}
    Acc. =  \frac{TP + TN}{TP + TN + FP + FN} 
\end{equation}

\item\textit{Fooling Rate} (FR) is the fraction of malicious samples misclassified as benign during the adversarial attack ($M'$) over the total number of malicious samples ($M$).

\begin{equation}
    FR =  \frac{M^{'}}{M} \times 100
\end{equation}

\end{itemize} 

\subsection{Experimental Results}
In this section, we will report the experimental evaluation of various proposed attacks and defence strategies under different scenarios. 

\subsubsection{Feature Engineering}

As discussed in the previous sub-section \ref{Feature Extraction}, our dataset consists of 197 features, each depicting individual permission that an application has requested during the installation. The datasets which contain a large feature vector will suffer from the curse of dimensionality and will have the following issue: (I) Higher training time and risk of overfitting in classification/detection models (II) Higher training time and large Q-table size of the RL agent. A Q-table with 197 features will have $2^{197}$ states and $198$ actions, which finally leads to  $2^{197} \times 198$ Q-table entries. It also makes the goal of extracting an optimal policy infeasible. To overcome this limitation, we perform feature selection to reduce the size of the feature vector, further substantially reducing the size of the Q-table.

Table \ref{featurereduction} shows the accuracy of various malware detection models constructed using different classification algorithms. When all the 197 permission features are considered for model construction, the highest accuracy of $93.81\%$ is achieved by RF followed by DT ($91.74\%$) and ET ($91.65\%$). SVM based model received the lowest accuracy of $85.42\%$ for the same. However, the current feature vector suffers from the curse of dimensionality for both constructing the malware detection model and the Q-table. To overcome this limitation, we perform feature selection using \textit{feature importance} to reduce the size of the feature vector for malware detection model.  It also substantially reduces the size of Q-table. Building the RF model with top $10$ (Appendix \ref{permissionrank}) features from \textit{feature importance} achieved the highest accuracy of $80.45\%$, and also the Q-table will consist of only $11,264$ entries. However, increasing the attributes to top $15$ features will increase the accuracy of the random forest model by a slight margin of $5\%$ but with a substantial increase in Q-table size with $5,242,88$ entries. Further selecting top $20$ and $25$ features will increase the Q-table entries to $22,020,096$ and $872,415,232$ respectively. It is also observed that training the RL agent with a more generalized detection model helps in finding better optimal policy.

\begin{table}[ht]
\caption{Top $10$ Android permission based on feature importance}
\begin{tabular}{|c|l|}
\hline
\textbf{Rank} & \multicolumn{1}{c|}{\textbf{Android Permission}}     \\ \hline
1             & android.permission.READ\_PHONE\_STATE                \\ \hline
2             & android.permission.READ\_SMS                         \\ \hline
3             & android.permission.SEND\_SMS                         \\ \hline
4             & android.permission.RECEIVE\_BOOT\_COMPLETED          \\ \hline
5             & android.permission.INSTALL\_PACKAGES                 \\ \hline
6             & android.permission.WRITE\_SMS                        \\ \hline
7             & android.permission.GET\_ACCOUNTS                     \\ \hline
8             & android.permission.RECEIVE\_SMS                      \\ \hline
9             & android.permission.ACCESS\_WIFI\_STATE               \\ \hline
10            & android.permission.ACCESS\_LOCATION\_EXTRA\_COMMANDS \\ \hline
\end{tabular}
\label{permissionrank}
\end{table}

\begin{table}[ht]
\caption{Accuracy of detection models after feature reduction}
\begin{center}
\begin{tabular}{|c|c|c|c|c|}
\hline
\multicolumn{1}{|l|}{}                                                  & \multicolumn{4}{c|}{\textbf{Accuracy}}                                                                                                                                                                                                                                            \\ \hline
\textbf{\begin{tabular}[c]{@{}c@{}}Classification\\ Model\end{tabular}} & \textbf{\begin{tabular}[c]{@{}c@{}}Top 5\\ Features\end{tabular}} & \textbf{\begin{tabular}[c]{@{}c@{}}Top 10\\ Features\end{tabular}} & \textbf{\begin{tabular}[c]{@{}c@{}}Top 15\\ Features\end{tabular}} & \textbf{\begin{tabular}[c]{@{}c@{}}All 197\\ Features\end{tabular}} \\ \hline
LR                                                                      & 75.98                                                             & 77.57                                                              & 81.09                                                              & 86.62                                                               \\ \hline
SVM                                                                     & 70.35                                                             & 76.93                                                              & 80.03                                                              & 85.42                                                               \\ \hline
DT                                                                      & 76.19                                                             & 80.45                                                              & 85.23                                                              & 91.74                                                               \\ \hline
RF                                                                      & 76.59                                                             & 80.95                                                              & 85.39                                                              & 93.81                                                               \\ \hline
AB                                                                      & 75.94                                                             & 77.57                                                              & 80.03                                                              & 85.76                                                               \\ \hline
GB                                                                      & 76.19                                                             & 80.42                                                              & 85.02                                                              & 91.65                                                               \\ \hline
ET                                                                      & 76.19                                                             & 80.45                                                              & 83.87                                                              & 88.09                                                               \\ \hline
DNN                                                                     & 76.47                                                             & 78.81                                                              & 82.72                                                              & 88.48                                                               \\ \hline
\end{tabular}
\end{center}
\label{featurereduction}
\end{table}

\subsubsection{Single-Policy Attack on Malware Detection Models}\label{para_a1q}

\begin{figure}[htbp!]
	\centering
	\includegraphics[width=\linewidth]{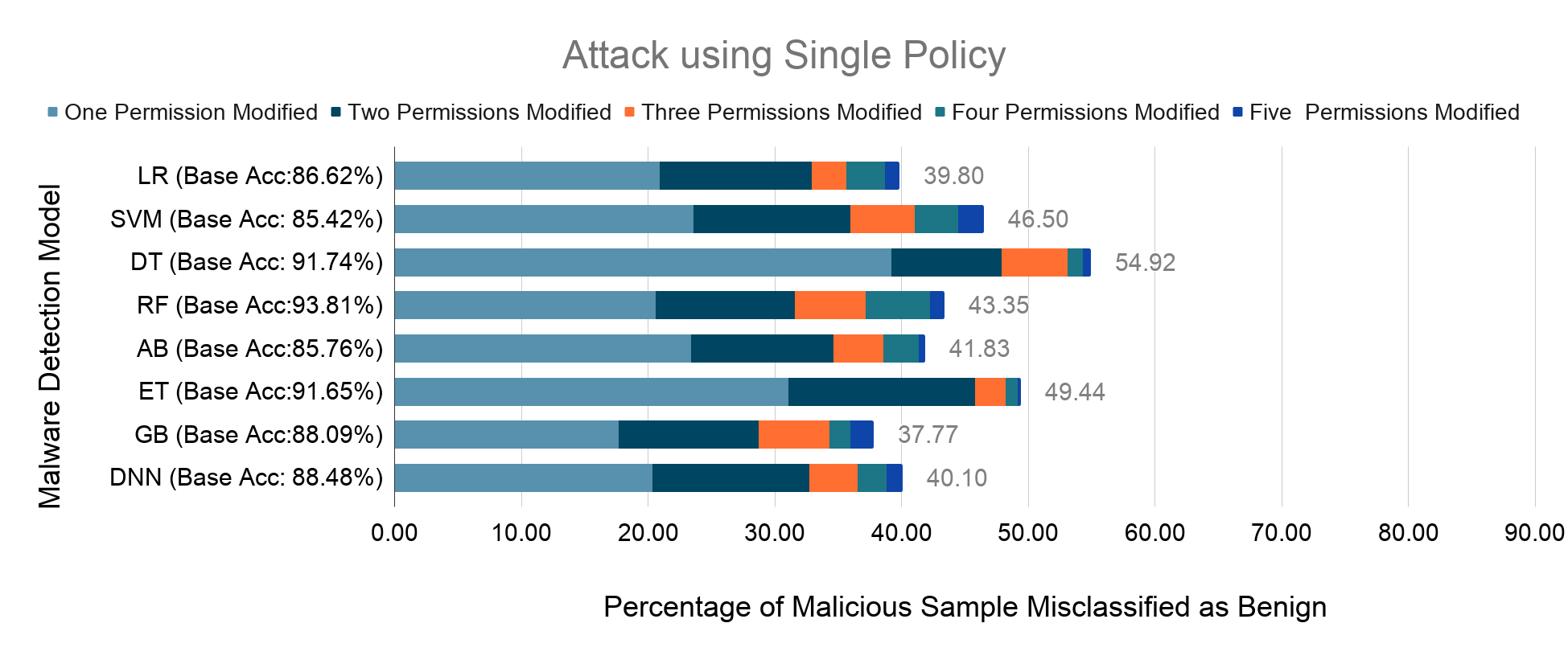}
	\centering
	\caption{Fooling rate achieved by SPA on different malware detection models}
	\label{A1Q}
\end{figure}

In SPA, an agent uses the optimal policy $\pi^*$ to modify the set of malware samples so that they are misclassified as benign. In order to evaluate the performance of our RL agent, we attack eight different malware detection models. Figure \ref{A1Q} shows the performance of the agent created using a Single Q-table against different malware detection models. In the plot, the vertical axis represents eight malware detection models and horizontal axis represents the percentage of malware samples successfully modified using a SPA. The vertical axis also contains baseline Android malware detection accuracy without any attack. RF model achieved the highest baseline detection accuracy of $93.81\%$ followed by the DT model with $91.74\%$.

The cost of an adversarial attack is calculated based on the number of permissions modified in the sample. Thus one permission modification attack allows maximum one permission to be modified using the optimal policy $\pi^*$. Similarly, two permission modification attack allow a maximum of two permission modifications using the same policy and so on. 

We started with one permission modification attack using single Q-table on all the eight detection models. The highest fooling rate of $39.19\%$ was achieved against the DT based Android malware detection model. It signifies that $1,959$ out of $5,721$ malware samples with only one permission modification has fooled the DT detection model and are misclassified as benign. The second highest fooling rate with one permission modification attack is attained for ET ($31.07\%$). The lowest fooling rate with one permission modification attack is obtained by GB ($17.66\%$), which is based on boosting mechanism. Using single policy for five permission modification attack allows a maximum of five permission modifications based on the optimal policy. The highest fooling rate with five permission modifications is achieved for the DT detection model ($54.92\%$), followed by the ET model ($49.44\%$). SPA achieved an average fooling rate of ($44.21\%$) among eight detection models, which is best in class.

\subsubsection{Multi-Policy Attack on Malware Detection Models}\label{para_anq}

\begin{figure}[htbp!]
	\centering
	\includegraphics[width=1.0\linewidth]{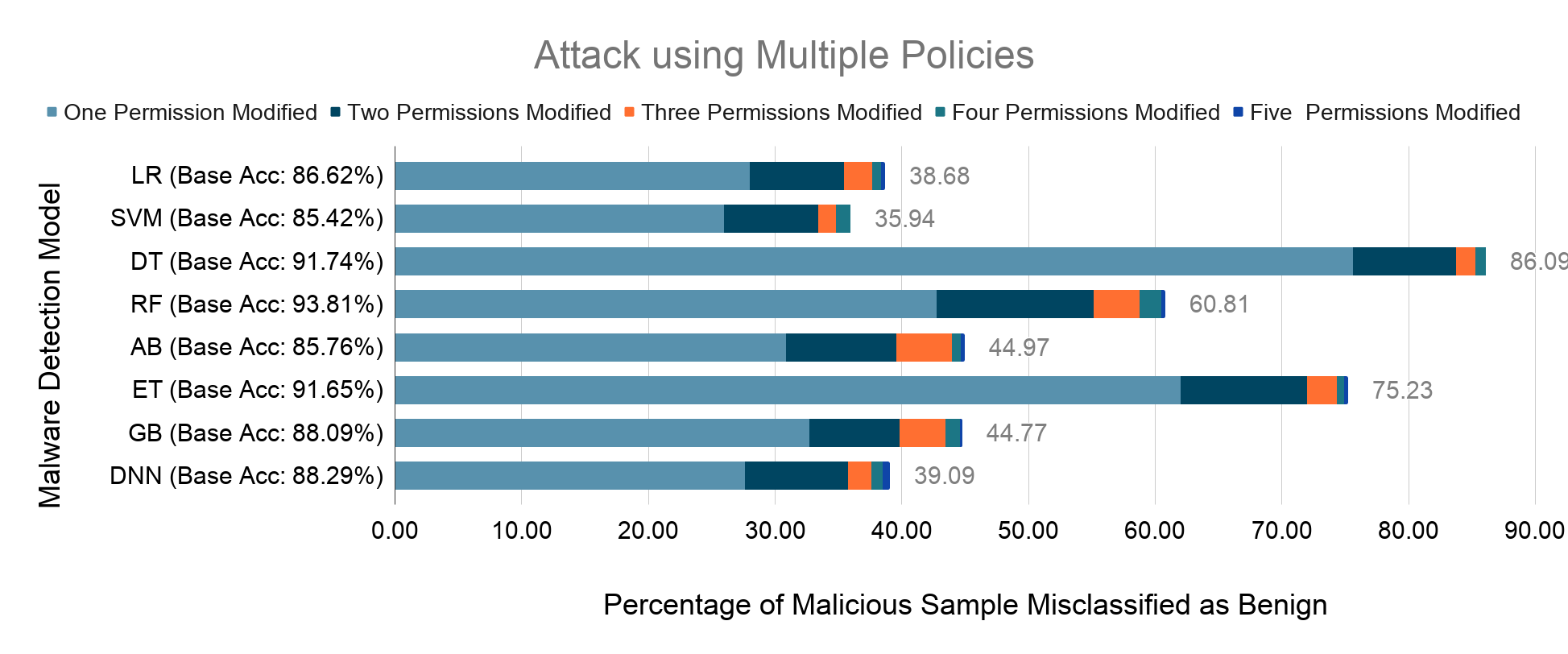}
	\centering
	\caption{Fooling rate achieved by MPA on different malware detection models}
	\label{AnQ}
\end{figure}

RL agent’s performance is improved when there is a large intersection between the features of the malware detection model and the RL agent. To overcome this limitation, we proposed a grey-box attack using several agents trained on different detection models. The strategy of using several detection models ensures a greater intersection of attributes irrespective of the detection model. Thus in an attack using multiple policies, an agent uses a set of optimal policies $\Pi$ obtained from multiple Q-tables to modify a malware sample to be misclassified as benign. Plot \ref{AnQ} shows the different malware detection models and their corresponding fooling rates. It is observed that the MPA outperforms the SPA as it achieves a better fooling rate against different malware detection models. The fooling rate with a maximum of five permission modification against the DT model is increased to $86.09\%$, which is a significant improvement over the single policy attack ($56.75\%$). A similar increase in the fooling rate is observed for the ET model ($75.23\%$) and RF model ($60.81\%$). The average fooling rate is increased to ($55\%$) for the eight Android malware detection models. As per our knowledge this the first investigation of the adversarial attacks on malware detection models in the grey-box scenario.

As the fooling rate against every malware detection model has improved, we can say that even with significant changes in the malware detection model, MPA can still attack with high accuracy. The multiple Q-table agents also perform better in scenarios where retraining the agent is not feasible, and the deployment can happen only once, as it is more adaptive to changes in the malware detection model. In the past, some researchers have proposed adversarial attacks similar to SPA for white-box scenario on Android malware detection models. Chen et al. designed three different variants of adversarial attacks namely week case, strong case and sophisticated case and modified the malicious Android applications to be misclassified as benign. Suciu et al. proposed an adversarial attack on detection models based on convolutional neural network and modified the executables with single step attacks. Although there is limited work published for adversial attacks in a white-box scenario. However, to the best of our knowledge, MPA is the first investigation of the adversarial attacks for grey box scenario.

\subsubsection{Defence against Single-Policy Attack}

After successfully increasing the fooling rate using multiple optimal policies obtained from multiple Q-tables, we design the defence strategy for the attack against the malware detection models. The above attack strategies demonstrated that adversaries can target the existing deployed models which acted as a motivating factor for developing  defensive mechanisms. In this  Kurakin et al. proposed that injecting adversarial samples while retraining the existing detection models significantly improved the robustness against known and zero-day attacks \cite{kurakin2016adversarial}.

\begin{figure}[htbp!]
	\centering
	\includegraphics[width=1.0\linewidth]{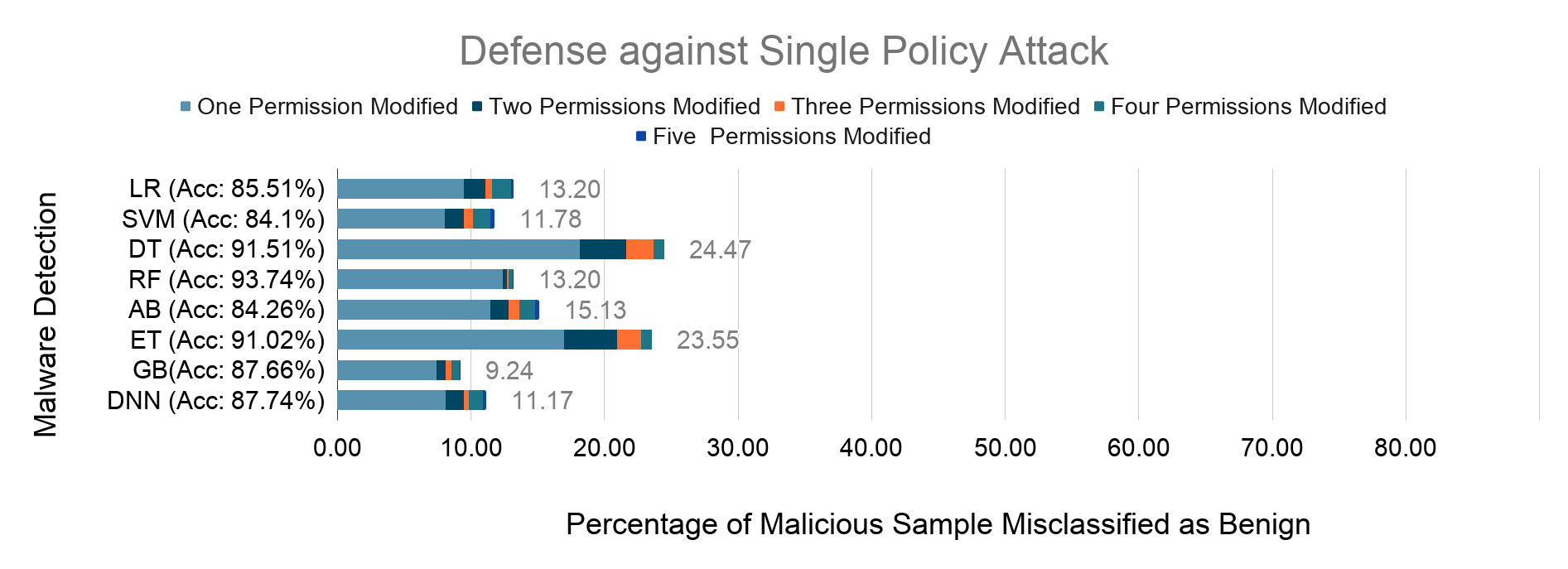}
	\centering
	\caption{Fooling rate achieved by SPA after the adversarial defence of detection models}
	\label{D1Q}
\end{figure}

Section \ref{para_a1q} and \ref{para_anq} shows that malware detection models are vulnerable to adversarial attacks. Also, as discussed in subsection \ref{defense_theory} that the process of injecting adversarial samples while retraining improves the security. Based on the figure \ref{AnQ}, it is observed that during an adversarial attack the RL agent modifies a substantial number of samples to be misclassified as benign. These benign-looking samples are stored and will be used to retrain the models to evaluate the model’s resilience against the zero-day attacks. The process consists of creating the training set with benign samples and selecting a percentage of new and old malware samples in order to have a class balanced dataset using random oversampling. Figure \ref{D1Q} shows the performance of the agent using the SPA against different malware detection models after retraining. In the figure, the vertical axis represents eight malware detection models and their corresponding baseline detection accuracy without any attack. Also, the horizontal axis represents the percentage of malware samples successfully modified using the SPA after retraining. It is observed that there is a slight drop in the model accuracy after retraining with new malware samples, but this is negligible compared to the increase in the robustness. From the figure, we can observe that the RF model achieves the highest baseline detection accuracy of $93.74\%$, which is slightly lower compared to before training accuracy of $93.81\%$. The baseline accuracy of the DT model also reduces from $91.74\%$ to $91.51\%$. 

The figure \ref{D1Q} shows a significant decrease in the fooling rate after retraining of detection models with correct labels. Comparing the retrained DT model with the original DT model, there is a considerable reduction in fooling rate from  $55.46\%$ to $24.46\%$. It is a substantial improvement which could be further increased by continuous retraining on new malware samples. Also, the fooling rates after retraining of ET and RF were reduced to $23.55\%$ and $13.30\%$ from earlier $49.44\%$ and $43.35\%$ respectively. The average fooling rate across all the eight detection models is reduced to one third from $44.21\%$ to $15.22\%$, which is a significant improvement. The drop in fooling rate also shows the limitation of single policy agents that they are not immune to changes in the malware detection system.

\subsubsection{Defence against Multi-Policy Attack}

\begin{figure}[htbp!]
	\centering
	\includegraphics[width=1.0\linewidth]{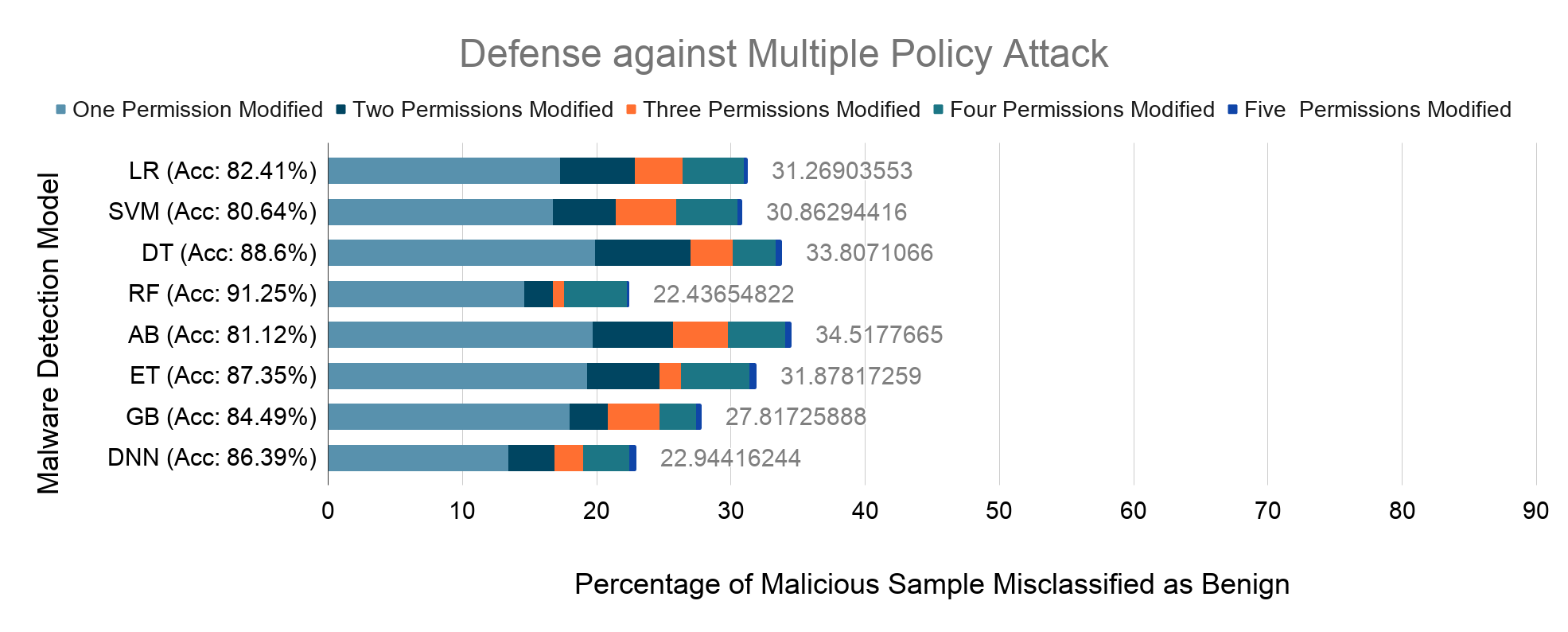}
	\centering
	\caption{Fooling rate achieved by MPA after the adversarial defence of detection models}
	\label{DnQ}
\end{figure}

We are following a similar retraining strategy for defense against the attack using MPA on Android malware detection system. The malware samples used for retraining are obtained from MPA but with correct labels. As the RL agent using multiple policies modifies a higher percentage of samples compared to the agent using a single policy, we observe a higher drop in the baseline accuracies mentioned on the vertical axis in figure \ref{DnQ}. From the figure, we can observe that the RF model achieved the highest baseline detection accuracy of $91.25\%$, which is slightly lower compared to before retraining accuracy of $93.81\%$. The horizontal axis represents the percentage of malware samples successfully modified using MPA after retraining.

Figure \ref{DnQ} shows a significant decrease in the fooling rate using RL agents with multiple policies. There is a reduction of $60.73\%$ in the fooling rate of DT from the retraining after MPA compared to SPA. The average fooling rate after retraining using multiple policy attack is $29.44\%$, which is approx half before the retraining. In general, the fooling rate can be further reduced by continuous retraining using new malware samples. However, the fooling rate achieved using multiple policy agents still outperforms single policy agent’s fooling rate. The presence of higher fooling rate even after adversarial training shows that multiple policy agents are more immune to changes in the malware detection system. Due to this fact they are preferred in scenarios where the malware detection models change frequently.

The figure \ref{D1Q} and \ref{DnQ} shows a drastic decrease in the fooling rate after one iteration of adversarial retraining. It also shows that the performance of the retrained models is nearly indistinguishable from baseline, whereas the robustness is drastically improved. While this experiment might have a few limitations, it provides evidence that this defense strategy could be used to improve robustness of the existing malware detection systems.

\section{Related Work}

Malware analysis and detection is a rat race between malware designers and the anti-malware community. Currently, anti-viruses mostly works on signature-based detection engines \cite{ye2017survey}  \cite{tam2017evolution} \cite{faruki2014android}. A signature is a unique value which indicates the presence of malicious code in the test application \cite{ye2017survey} \cite{sharma2014evolution}. These signatures are stored in the anti-virus database and compared with the test application during anti-virus scanning. If any signature from the database matches the test application, then it is declared as malicious otherwise not \cite{ye2017survey}. The signature-based mechanism is vastly human-driven, time-consuming and does not provide security against zero-day attacks \cite{ye2017survey}. Heuristic and behaviour-based detection mechanisms often complement the signature engines \cite{ye2017survey}  \cite{tam2017evolution}. In heuristic-based detection, rules are developed to separate malicious and benign applications. Typically domain experts draft these rules which can also detect zero-day attacks. Developing generic rules to detect malware samples is again human-driven, error-prone and often hard to achieve without increasing the false positive rate \cite{ye2017survey}. Behaviour-based detection mechanism check for run-time triggers that can capture properties of the malware applications. These signature, heuristic, and behaviour based mechanisms are vastly human dependent, time-consuming and not scalable and thus cannot detect next-generation polymorphic/metamorphic malware \cite{ye2017survey}.

During the past few years, the anti-malware research community has investigated the  development of Android malware detection systems based on machine learning and deep learning which has shown promising results. Developing these solutions is a two-step process: 1) Feature Engineering, and 2) Classification/Clustering. In the first phase of feature engineering,  extraction of features are performed using static, dynamic or hybrid analysis \cite{ye2017survey}  \cite{tam2017evolution}. If the features are extracted without executing an Android application, then it is known as static analysis. In this, Arp et al. in 2014 extracted roughly $545,000$ features by static analysis from the Android applications and constructed Android malware detection model using support vector machine \cite{arp2014drebin}. On the other hand, when features are extracted by executing an Android application inside a sandbox or controlled environment, then it is called dynamic analysis. Dash et al. in DroidScribe extracted features like network addresses, file access, binder methods etc. and proposed a malware detection model using support vector machine \cite{dash2016droidscribe}. In the hybrid analysis, features extracted using both static and dynamic analysis are combined to develop the malware detection models. SAMADroid proposed by Arshad et al. used six static features (permission, intent etc.), one dynamic feature (system call) and four classification algorithms (support vector machine, decision tree, etc.) to construct the malware detection system \cite{arshad2018samadroid}.

In the second phase of Classification/Clustering, various machine learning and deep learning algorithms can be used to construct the classification models. Wu et al. proposed Droidmat which extracted permissions, intents, API calls etc. from Android applications. They performed feature reduction using single value decomposition followed by building Android malware detection model based on the k-nearest neighbors algorithm and achieved an accuracy of $97.87\%$ \cite{wu2012droidmat}. Droid-Sec by Yuan et al. extracted more than $200$ features from Android apps and constructed malware detection models using SVM, LR, DNN etc. They achieved the highest detection accuracy of  $96.5\%$ with DNN based model \cite{yuan2014droid}. Yerima et al. constructed a malware detection model based on the Bayesian approach and attained an accuracy of $92.1\%$ \cite{yerima2013new}. Lindorfer et al. in ANDRUBIS downloaded more than $1,000,000$ apps and found that malicious applications tend to request more permissions than benign applications \cite{lindorfer2014andrubis}. Sewak et al. used deep learning based autoencoders and deep neural network for developing malware detection models and achieved an accuracy of $99.21\%$ \cite{sewak2018comparison}. Rathore et al. proposed Android malware models based on clustering and classification and achieved the highest accuracy of $95.82\%$ with random forest \cite{rathore2018android}. Appice et al. proposed MuViDA, which collected features like permission, intent, API calls from Android applications. They developed additional features with clustering and classification algorithms and achieved the highest AUC of $96.6\%$ \cite{appiceclustering}.

With the rise of machine learning models, evasion attacks have gained popularity. One such attack is the mimicry attack that had successfully reduced the detection model accuracy by injecting beingness into malware samples to misclassify them. DNN models are also susceptible to such attacks, Grosse et al., Kolosnjaji et al. and others demonstrated this by successfully misleading a neural network using carefully crafted samples \cite{paudice2018label} \cite{grosse2017adversarial} \cite{chen2018automated}. Their idea involved using a genetic algorithm to add perturbations in the malware file to evade the detection system. AL can also be used to perform grey-box attacks. Battista Biggio showed that even with limited knowledge of the target system, classification algorithms can be evaded with very high probability \cite{biggio2013evasion}. With the rise of adversarial attacks, adversarial learning has also gained popularity among different machine learning domains like computer vision, spam detection etc. Ian J. Goodfellow \cite{goodfellow2014explaining}, Florian Tramer \cite{tramer2017ensemble}, and Aleksander Madry \cite{madry2017towards} have shown the advantage of using adversarial learning to improve the robustness of neural network models. Adversarial learning also improves resilience against a single step attack as shown by Alexey Kurakin \cite{kurakin2016adversarial}. Based on above discussions, it can be inferred that using AL can improve the robustness of existing models.

\section{Discussion and Conclusion}

Today smartphones and Android OS have become an integral part of our society. Traditionally, Android malware detection mechanisms like the signature based, heuristic based, behavioural based etc.  are not able to cope up against the current attacks developed by malware designers. Thus recently, machine learning and deep learning based malware detection models have attracted the interest of the anti-malware community. However, these models perform poorly against adversarial attacks which could threaten the complete malware detection ecosystem.

To develop a robust Android malware detection system against adversarial attacks, we first build eight different malware detection models. The models are constructed using a variety of classification techniques like traditional algorithms (LR, SVM, and DT), bagging algorithms (RF, ET), boosting algorithms (AB, GB) and DNN. The highest malware detection accuracy is achieved using RF ($93.81\%$), followed by the DT model ($91.94\%$). Also, all the other detection models have attained more than $85\%$ accuracy. However, these models are susceptible to adversarial attacks designed by the malware architects based on the knowledge about the system.  It can be visualized as a min-max game where adversary wants to maximize the fooling rate with minimum modifications and without any functional or behavioural disturbance to the Android application.
For the purpose, we created new variants of malware using RL, which will be misclassified as benign by the existing Android malware detection models.

We also proposed a novel single policy attack for the white-box setting where an adversary has complete knowledge about the detection system. We design a reinforcement agent which performs adversarial attack using a policy obtained from a single Q-table. The attack achieves an average fooling rate of $44.28\%$ across all the eight detection models with maximum five modifications. The attack also achieves the highest fooling rate against the DT model ($54.92\%$) whereas the lowest fooling rate is obtained for GB ($37.77\%$) with similar setting.  Overall, the experimental result signifies that single policy attack can successfully evade malware detection models and accomplish high fooling rate even with limited modifications.  

We also develop a state-of-the-art adversarial attack, namely multi-policy attack for grey-box setting where the attacker has no knowledge about the model architecture and classification algorithm. The multi-policy attack achieve the highest fooling rate for DT model ($86.09\%$) followed by ET model ($75.23\%$) with a maximum of five modifications. The average fooling rate is increased to $53.20\%$, which is higher than the single policy attack even with limited information.

Finally, we propose defense against adversarial attacks based on the single policy and multi-policy attack strategies. With adversarial retraining, the average fooling rate against the single policy attack is reduced by threefold to $15.22\%$ and twofold for the multi-policy attack to $29.44\%$ i.e. it can now effectively detect variants (metamorphic) of malware. The experimental analysis shows our proposed Android malware detection system using reinforcement learning is more robust against adversarial attacks.

In this work, we have used Android permission as a feature and used Q learning for designing adversarial attacks on Android malware detection models. However, many researchers have proposed malware detection models built using features from static analysis (Intent, API calls, etc.) and dynamic analysis (system calls). In future, we will explore fooling Android malware detection models based on other features. We are also planning to design adversarial attack based on other reinforcement learning techniques like deep q-learning, actor-critic algorithm, proximal policy optimization etc.

\bibliographystyle{plain}      
\bibliography{main.bib}   

\end{document}